\documentclass[12pt,halfline,a4paper]{ouparticle}
\usepackage{graphicx}
\usepackage{multirow}
\usepackage{caption}
\usepackage{lscape}
\usepackage{adjustbox}
\usepackage[detect-all]{siunitx}
\usepackage{siunitx}
\usepackage{longtable}
\usepackage{lscape}
\usepackage{float}
\usepackage{amsmath,amssymb,amsfonts}
\usepackage{amsthm}
\usepackage{array}
\usepackage{cite}
\usepackage{tikz}
\usepackage{mathrsfs}
\usepackage[title]{appendix}
\usepackage{graphicx}
\usepackage[table]{xcolor}
\usepackage{float}
\usepackage{textcomp}
\usepackage{manyfoot}
\usepackage{booktabs}
\usepackage{algorithm}
\usepackage{algorithmicx}
\usepackage{algpseudocode}
\usepackage{listings}
\usepackage{xcolor}
\usepackage{float} 
\usepackage[font=small,labelfont=bf,justification=centering]{caption}
\usepackage{subcaption}
\newtheorem{theorem}{Theorem}[section]

\begin{document}

\title{Edge Irregularity Strength: A Complementary Descriptor to Topological Indices in QSPR and QSAR Studies}

\author{%
\name{U. Vijaya Chandra Kumar}
\address{Department of Mathematics, School of Applied Sciences \\
REVA University, Bengaluru, Karnataka, India}
\email{uvijaychandra.kumar@reva.edu.in}
\and
\name{H.M. Nagesh$^{*}$}
\address{Department of Science and Humanities \\
PES University, Bengaluru, Karnataka, India}
\email{nageshhm@pes.edu}
\and\name{Narahari N}
\address{Department of Mathematics \\
University College of Science, Tumkur University \\
Tumakuru, Karnataka, India}
\email{narahari@tumkuruniversity.ac.in}
$^{*}$Corresponding author 
}
\newpage

\abstract{In chemical graph theory, topological indices are widely used as numerical descriptors for establishing quantitative structure–property relationships (QSPR) and quantitative structure–activity relationships (QSAR). These indices successfully correlate molecular structure with various physicochemical and biological properties. In addition to these methods, the concept of edge irregularity strength, a graph labeling measure, offers another perspective for representing structural characteristics. In this context, the edge irregularity strength concept provides a systematic way of assigning numerical labels to atoms based on specific rules. In this work, we explore the chemical applicability of the edge irregularity strength and demonstrate that it can also serve as a predictive tool for physicochemical properties, similar to topological indices. The findings show that the edge irregularity strength captures molecular features and complements existing approaches to structure–property analysis in chemical graph theory.}

\date{}


\maketitle

\section{Introduction} \label{sec:Intr}

Let \( G \) be a connected, simple, undirected graph with vertex set \( V(G) \) and edge set \( E(G) \). 
A \emph{labeling} is a function that assigns a set of numbers—typically positive integers, called \emph{labels}—to elements of the graph. 
If the function is defined on the vertex set, it is referred to as a \emph{vertex labeling}, whereas if it is defined on the edge set, it is called an \emph{edge labeling}. 
When the labeling is defined on both \( V(G) \cup E(G) \), it is known as a \emph{total labeling}. For an edge \( k \)-labeling \( f: E(G) \to \{1, 2, \ldots, k\} \), the \emph{weight} of a vertex \( x \in V(G) \) is given by $w(x) = \sum f(xy)$, where the summation extends over all edges incident with \( x \).

Chartrand et al. \cite{1} defined irregular labeling for a graph $G$ as an assignment of labels from the set of natural numbers to the edges of $G$ such that the sum of the labels assigned to the edges of each vertex is different. The minimum value of the largest label of an edge over all existing irregular labeling is called the \emph{irregularity strength} of $G$ and is denoted by $s(G)$. Determining $s(G)$ can be challenging, even for graphs with simple structures \cite{1}. For instance, as illustrated in Figure 1, the irregularity strength of the Petersen graph is 5. 

\begin{figure}[hbt!]
	\centering
	\includegraphics[width=80mm]{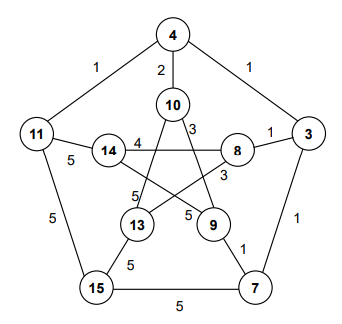}
	\textbf{\caption{Irregularity strength of the Petersen graph}} 
\end{figure} 

Ahmad et al. \cite{2} introduced the concept of edge irregular $k$-labeling of graphs. A vertex labeling $\phi:V(G) \rightarrow \{1, 2,\ldots,k\}$ is called \emph{$k$-labeling}. The \emph{weight} of an edge $xy$ in $G$, written $w_{\phi}(xy)$, is the sum of the labels of end vertices $x$ and $y$, i.e., $w_{\phi}(xy)=\phi(x)+\phi(y)$. A vertex $k$-labeling of a graph $G$ is defined to be an edge irregular $k$-labeling of the graph $G$ if for every two different edges $e$ and $f$, $w_{\phi}(e) \neq w_{\phi}(f)$. The minimum $k$ for which the graph $G$ has an edge irregular $k$-labeling is called the \emph{edge irregularity strength} of $G$, denoted by $es(G)$. It is obvious that $s(G)$ is an edge labeling of a graph $G$ in which distinct vertices have distinct weights, and $es(G)$ is  a vertex labeling of a graph $G$ in which any two distinct edges have distinct weights. 

Figure 2 shows a graph \(G\) together with its edge irregularity strength \(es(G) = 9\).

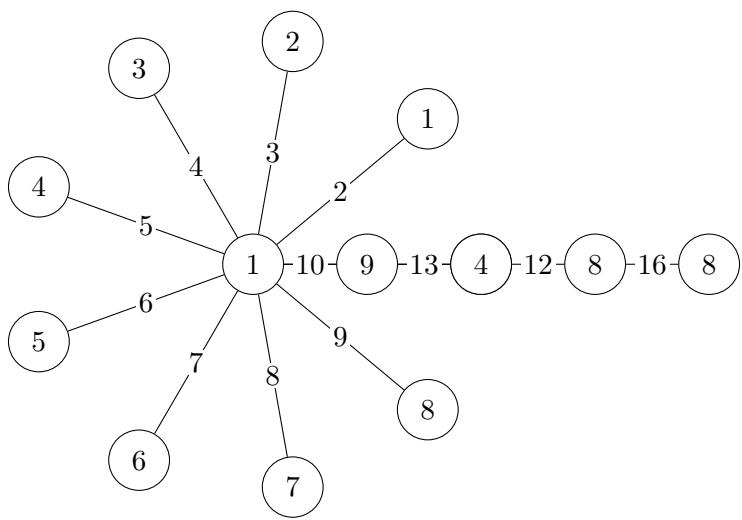
\begin{figure}[htbp]
\centering
\begin{tikzpicture}[
    scale=1.0, 
    vertex/.style={circle, draw=black, fill=white, inner sep=2pt, minimum size=8mm, font=\small},
    edge label/.style={midway, fill=white, font=\small, inner sep=1pt}
]

% Star parameters
\def\nStar{9}  
\def\rStar{3}  
\def\starLabels{{1,1,2,3,4,5,6,7,8}} 

% Path parameters
\def\pathLabels{{9,4,8,8}} 
\def\nPath{4} 
\def\dx{1.5}    

% Central vertex
\node[vertex] (c) at (0,0) {1};

\foreach \i in {1,...,\nStar} {
    \pgfmathsetmacro{\angle}{360/\nStar * (\i - 1)}
    \pgfmathtruncatemacro{\lab}{\starLabels[\i - 1]}
    \node[vertex] (s\i) at ({\rStar*cos(\angle)}, {\rStar*sin(\angle)}) {\lab};
    \draw (c) -- (s\i) node[edge label] {\pgfmathtruncatemacro{\w}{\lab+1}\w};
}

\foreach \j in {1,...,\nPath} {
    \pgfmathsetmacro{\x}{\dx*\j}
    \pgfmathtruncatemacro{\lab}{\pathLabels[\j - 1]}
    \node[vertex] (p\j) at (\x,0) {\lab};
    
    \ifnum\j=1
        \draw (c) -- (p1) node[edge label] {10};
    \else
        \pgfmathtruncatemacro{\prev}{\pathLabels[\j - 2]}
        \pgfmathtruncatemacro{\sum}{\prev + \lab}
        \draw (p\the\numexpr\j-1) -- (p\j) node[edge label] {\sum};
    \fi
}

\end{tikzpicture}
\caption{A graph $G$ and its $es(G)=9$}
\end{figure}

For further details on the edge irregularity strength of graphs, readers are referred to \cite{3,4,5,6,7,8,9}.

Topological indices (TI’s) are molecular descriptors that link chemical structure and important physicochemical and biological activity \cite{10, 11}. These topological descriptors are part of a collection of theoretical tools for describing these molecules’ structural properties. As a result, TI’s and their evolution have received much attention. Quantitative Structure–Property Relationship (QSPR) modeling is a powerful computational approach used to predict the physicochemical properties and biological activities of pharmaceutical compounds based on molecular descriptor analysis \cite{12}. In this context, topological indices (TIs) have emerged as essential mathematical tools that convert molecular structural information into numerical descriptors, enabling accurate property–activity correlations and predictions \cite{13,14}. 

\subsection{Motivation and Novelty}
Traditional topological indices have long been valuable for linking molecular structure to physical, chemical, and biological properties. However, as molecular systems increase in complexity, these conventional indices may fail to capture certain structural irregularities. The concept of edge irregularity strength, originating from graph labeling theory, offers a novel approach by assigning numerical labels to atoms under specific constraints, thereby quantifying the degree of irregularity within a molecular graph. This measure provides a fresh perspective on molecular topology and holds promise as a complementary molecular descriptor. Incorporating edge irregularity strength into QSPR and QSAR modeling can enhance predictive accuracy and deepen our understanding of molecular structure beyond what traditional topological indices offer.

\section{Main results}
The following theorem in \cite{2} establishes the lower bound for the edge irregularity strength of a graph $G$. 
\begin{theorem}
Let $G=(V,E)$ be a simple graph with maximum degree $\Delta(G)$. Then 
\begin{center} 
$es(G) \geq \max \{\lceil \frac{|E(G)|+1}{2} \rceil, \Delta(G)\}.$
\end{center} % TECH EDITOR : max > \max
\end{theorem}
The important findings in this paper are proved using Theorem 2.1.

\subsection{Chemical relevance of edge irregularity strength}
Topological indices assign numerical values to molecular structures to quantify physicochemical properties. In this context, we now show that the edge irregularity strength also satisfies the same set of desirable criteria, thereby confirming it as a meaningful descriptor. One such attribute is the ability to predict a molecule’s physicochemical properties. Evaluating effectiveness in modeling physicochemical properties involves linking structural features with experimental data across a benchmark set of compounds. If the correlation coefficient $R \geq 0.8$, the measure is included in the regression analysis \cite{15}. To explore the chemical significance of the edge irregularity strength, we examine the following relation. 

\begin{equation}\label{eq:linear-model}
Y = a\,X + b
\end{equation}
Here, 
\begin{align*}
Y &\text{ is the dependent property},\\
X &\text{ is the edge irregularity strength},\\
a &\text{ is the slope of the regression line},\\
b &\text{ is the intercept}.
\end{align*}

Throughout this work, unless otherwise specified, a vertex represents an atom, an edge corresponds to a bond, and any graph refers to the molecular graph of a benzenoid hydrocarbon (BH). We determine the edge irregularity strength of each benzenoid hydrocarbon by the procedure outlined below. 

To illustrate the method, we then work through a specific example. Although Theorem~2.1 yields a lower bound on the edge irregularity strength of a graph, our analysis requires the exact value of this parameter. Let $G$ be the molecular graph of Naphthalene. Then the number of edges is $|E(G)|=11$ and the maximum degree is $\Delta(G)=3$. Therefore, by Theorem 2.1, the edge irregularity strength of $G$ is 6, calculated as 
\[
\mathrm{es}(G)
\;=\;
\max\!\Bigl\{\bigl\lceil\tfrac{|E(G)|+1}{2}\bigr\rceil,\;\Delta(G)\Bigr\}
\;=\;
\max\{\lceil\tfrac{11+1}{2}\rceil,\;3\}
\;=\;
\max\{6,3\}
\;=\;
6,
\] as shown in Figure 3, where each vertex is labeled in the form $A:B$, where $A$ represents the vertex number $(1, 2, 3, \ldots)$ and $B$ denotes the vertex label assigned according to the definition of edge irregularity strength. The edge weights are calculated as the sum of the vertex labels of the two vertices connected by that edge.
   
\begin{figure}[hbt!]
	\centering
	\includegraphics[width=90mm]{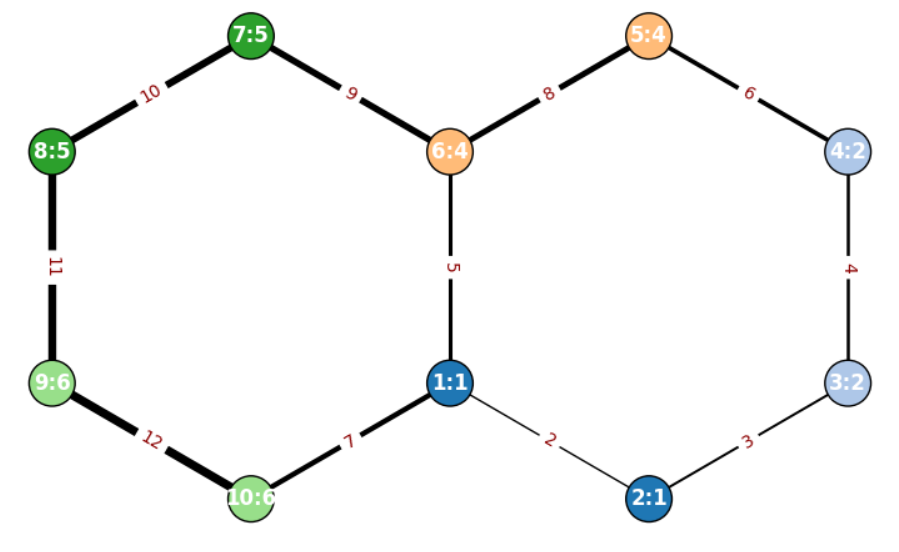}
	\textbf{\caption{Edge irregularity strength of Naphthalene}} 
\end{figure} 

The molecular graphs of the benzenoid hydrocarbons (BHs) under study are shown in Figure~4. Experimental properties—including boiling point (BP), $\pi$‐electron energy ($E_{\pi}$), molecular weight (MW), polarizability (PO), molar volume (MV), molar refractivity (MR), XLogP3, heavy atom count (HAC) and complexity (C)—were obtained from references~\cite{16}. 

\newpage

Table~1 presents these experimental values together with the computed edge irregularity strength for each molecular graph of a benzenoid hydrocarbon (BH). Table~2 shows the linear regression models relating the edge irregularity strength to the physicochemical properties of benzenoid hydrocarbons.

\vspace{5mm}

\begin{figure}[hbt!]
	\centering
	\includegraphics[width=150mm]{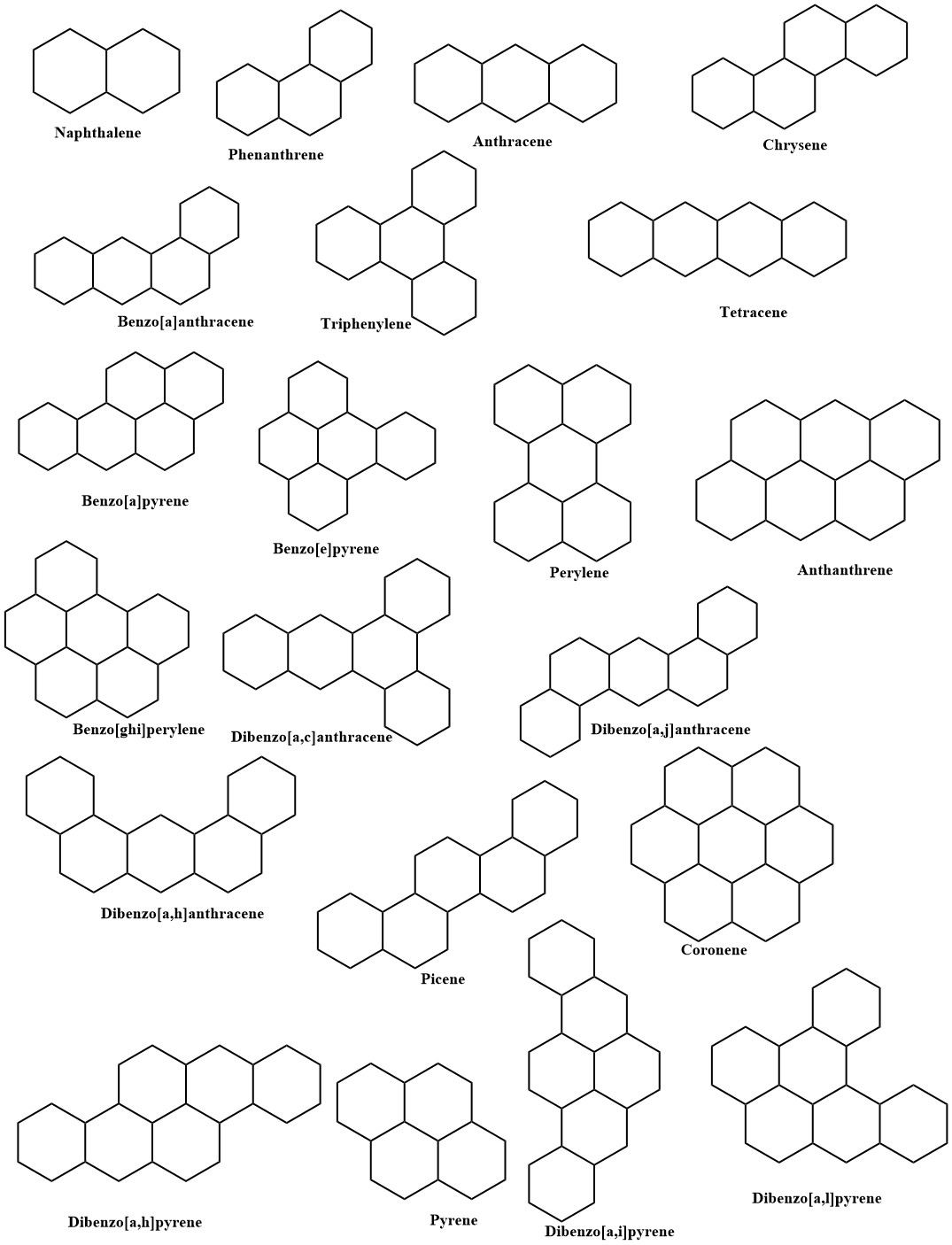}
	\textbf{\caption{Molecular graph of benzenoid hydrocarbons}} 
\end{figure} 

\newpage

\begin{table*}[htbp]
  \centering
  \resizebox{\textwidth}{!}{%
  \sisetup{
    table-format = 3.1,
    table-number-alignment = center,
  }
  \begin{tabular}{
      l
      S[table-format=2.0]      
      S[table-format=3.1]
      S[table-format=2.3]
      S[table-format=3.2]
      S[table-format=3.1]
      S[table-format=3.1]
      S[table-format=2.1]
      S[table-format=1.1]
      S[table-format=2.0]
      S[table-format=3.1]
  }
    \toprule
    {Molecules} & {Edge Irregularity Strength} & {BP} & {$E_{\pi}$} & {MW} & {PO} & {MV} & {MR} & {XLogP3} & {HAC} & {C} \\
    \midrule
    Naphthalene & 6   & 221.5 & 13.683 & 128.17 & 17.5 & 123.5 & 44.1 & 3.3 & 10 & 80.6 \\
    Phenanthrene & 9  & 337.4 & 19.448 & 178.23 & 24.6 & 157.7 & 61.9 & 4.5 & 14 & 174 \\
    Anthracene & 9  & 337.4 & 19.314 & 178.23 & 24.6 & 157.7 & 61.9 & 4.4 & 14 & 154 \\
    Chrysene & 11  & 448 & 25.192 & 228.3 & 31.6 & 191.8 & 79.8 & 5.7 & 18 & 264 \\
    Benzo[a]anthracene & 11  & 436.7 & 25.101 & 228.3 & 31.6 & 191.8 & 79.8 & 5.8 & 18 & 294 \\
    Triphenylene & 11  & 425 & 25.275 & 228.3 & 31.6 & 191.8 & 79.8 & 4.9 & 18 & 217 \\
    Tetracene & 11 & 436.7 & 25.188 & 228.3 & 31.6 & 191.8 & 79.8 & 5.9 & 18 & 236 \\
    Benzo[a]pyrene & 13  & 495 & 28.222 & 252.3 & 35.8 & 196.1 & 90.3 & 6.0 & 20 & 372 \\
    Benzo[e]pyrene & 13  & 467.5 & 28.336 & 252.3 & 35.8 & 196.1 & 90.3 & 6.4 & 20 & 336 \\
    Perylene & 13  & 467.5 & 28.245 & 252.3 & 35.8 & 196.1 & 90.3 & 5.8 & 20 & 304 \\
    Anthanthrene & 14  & 497.1 & 31.253 & 276.3 & 40.0 & 200.4 & 100.8 & 6.7 & 22 & 411 \\
    Benzo[ghi]perylene & 14  & 501 & 31.425 & 276.3 & 40.0 & 200.4 & 100.8 & 6.6 & 22 & 411 \\
    Dibenzo[a,c]anthracene & 14  & 518 & 30.942 & 278.3 & 38.7 & 225.9 & 97.6 & 6.7 & 22 & 361 \\
    Dibenzo[a,j]anthracene & 14  & 524.7 & 30.880 & 278.3 & 38.7 & 225.9 & 97.6 & 6.5 & 22 & 363 \\
    Dibenzo[a,h]anthracene & 14 & 524.7 & 30.881 & 278.3 & 38.7 & 225.9 & 97.6 & 6.5 & 22 & 361 \\
    Picene & 14  & 519 & 30.943 & 278.3 & 38.7 & 225.9 & 97.6 & 7.0 & 22 & 361 \\
    Coronene & 16  & 525.6 & 34.572 & 300.4 & 44.1 & 204.7 & 111.4 & 7.2 & 24 & 376 \\
    Dibenzo[a,h]pyrene & 15  & 552.3 & 33.928 & 302.4 & 42.9 & 230.2 & 108.1 & 7.0 & 24 & 436 \\
    Dibenzo[a,i]pyrene & 15  & 552.3 & 33.954 & 302.4 & 42.9 & 230.2 & 108.1 & 7.3 & 24 & 436 \\
    Dibenzo[a,l]pyrene & 15  & 552.3 & 34.031 & 302.4 & 42.9 & 230.2 & 108.1 & 7.2 & 24 & 480 \\
    Pyrene & 10  & 404 & 22.506 & 202.25 & 28.7 & 162.0 & 72.5 & 4.9 & 16 & 217 \\
    \bottomrule
  \end{tabular}}
  \caption{Edge irregularity strength values and experimental properties for benzenoid hydrocarbons.}
  \label{tab:benzenoid-properties}
\end{table*}

\begin{table*}[htbp]
\centering
\resizebox{\textwidth}{!}{%
\begin{tabular}{lccccccc}
\toprule
\textbf{Variable} & \textbf{Slope} & \textbf{Intercept} & \textbf{$R$} & \textbf{$R^2$} & \textbf{SE} & \textbf{F} & \textbf{SF (p-value)} \\
\midrule
BP      & 32.4152 &  59.5680 &  0.9714 &  0.9437 &  1.8168 &   318.3280 &  2.51$\times$10$^{-13}$ \\
$E_{\pi}$ &  2.1971 &   0.3656 &  0.9946 &  0.9892 &  0.0527 &  1738.4184 &  3.81$\times$10$^{-20}$ \\
MW      & 18.5868 &  17.1739 &  0.9920 &  0.9840 &  0.5432 &  1170.7772 &  1.55$\times$10$^{-18}$ \\
PO      &  2.7684 &   0.5469 &  0.9949 &  0.9899 &  0.0641 &  1866.7432 &  1.95$\times$10$^{-20}$ \\
MV      & 10.4545 &  67.4778 &  0.9170 &  0.8408 &  1.0436 &   100.3568 &  5.11$\times$10$^{-9}$ \\
MR      &  6.9817 &   1.3802 &  0.9949 &  0.9899 &  0.1621 &  1856.0632 &  2.06$\times$10$^{-20}$ \\
XLogP3  &  0.4146 &   0.8412 &  0.9696 &  0.9402 &  0.0240 &   298.8263 &  4.43$\times$10$^{-13}$ \\
HAC     &  1.5000 &   1.0000 &  0.9931 &  0.9862 &  0.0407 &  1359.8571 &  3.82$\times$10$^{-19}$ \\
C       & 39.5424 & -176.9295 &  0.9503 &  0.9030 &  2.9730 &   176.9060 &  4.47$\times$10$^{-11}$ \\
\bottomrule
\end{tabular}}
\caption{Linear regression models relating the edge irregularity strength to the physicochemical properties of benzenoid hydrocarbons}
\label{tab:regression-results}
\end{table*}

\noindent
The results presented in Table~\ref{tab:regression-results} show that all physicochemical properties of benzenoid hydrocarbons exhibit strong linear relationships with edge irregularity strength. All regression models demonstrate high correlation coefficients, with statistically significant p-values. Among the models, $E_{\pi}$, PO, and MR display the best fits, characterized by the highest correlation coefficients ($R^2 \approx 0.99$). Other variables, including BP, MW, XLogP3, and HAC, also show strong predictive capability with $R^2$ values in the range of 0.94--0.99. In comparison, MV and C exhibit slightly lower yet still substantial coefficients of determination ($R^2 = 0.84$ and $0.90$, respectively). Overall, these findings indicate that the physicochemical properties investigated are significantly correlated with edge irregularity strength. \newpage

The results indicate that the edge irregularity strength effectively captures structural variations influencing the physicochemical properties of benzenoid hydrocarbons. In particular, \(E_{\pi}\), PO, and MR are predicted with high accuracy (\(R^2 > 0.98\)), confirming the strong predictive capability of the edge irregularity strength.

Figures 5-13 show the scatter plots of the linear model relating the edge irregularity strength to the physicochemical properties of benzenoid hydrocarbons. To simplify our notation, we write the edge irregularity strength as $es$.

\vspace{3mm}
\begin{figure}[hbt!]
	\centering
	\includegraphics[width=120mm]{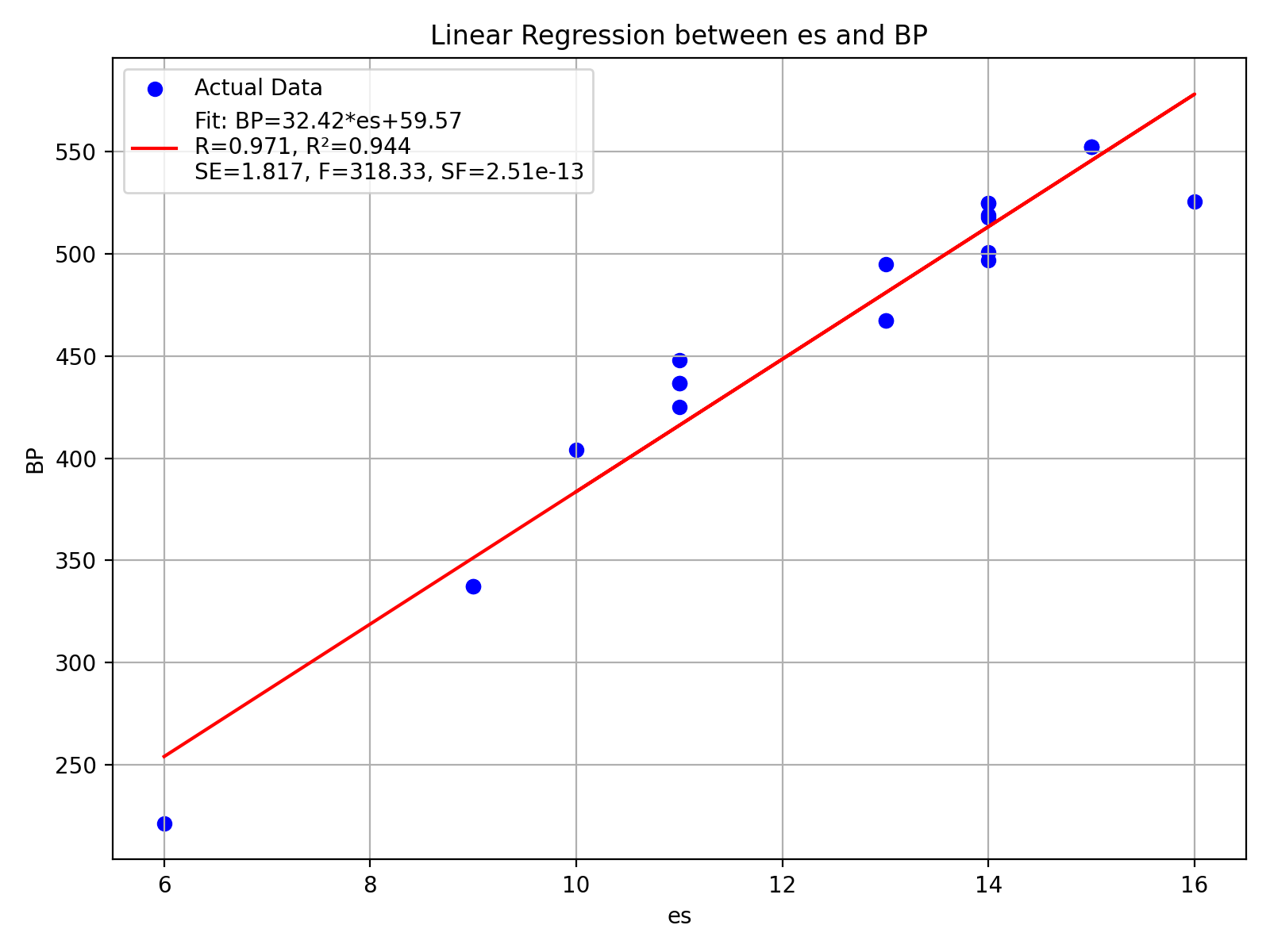}
	\caption{Scatter plot of the linear model between the edge irregularity strength and the boiling point property of benzenoid hydrocarbons}
\end{figure} 

\begin{figure}[hbt!]
	\centering
	\includegraphics[width=120mm]{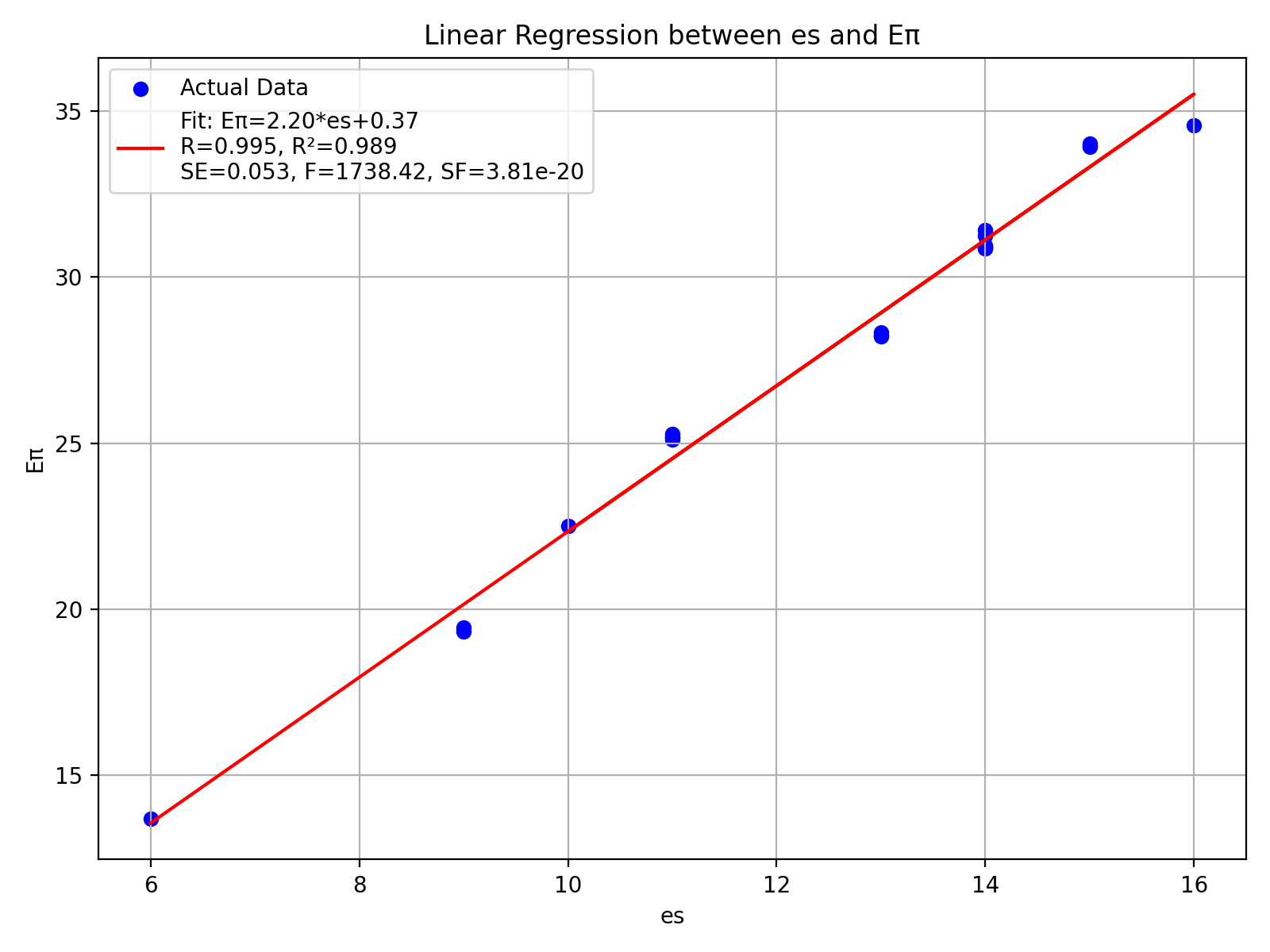}
	\caption{Scatter plot of the linear model between the edge irregularity strength and the $\pi-$electron energy property of benzenoid hydrocarbons}
\end{figure} 

\newpage
\begin{figure}[hbt!]
	\centering
	\includegraphics[width=120mm]{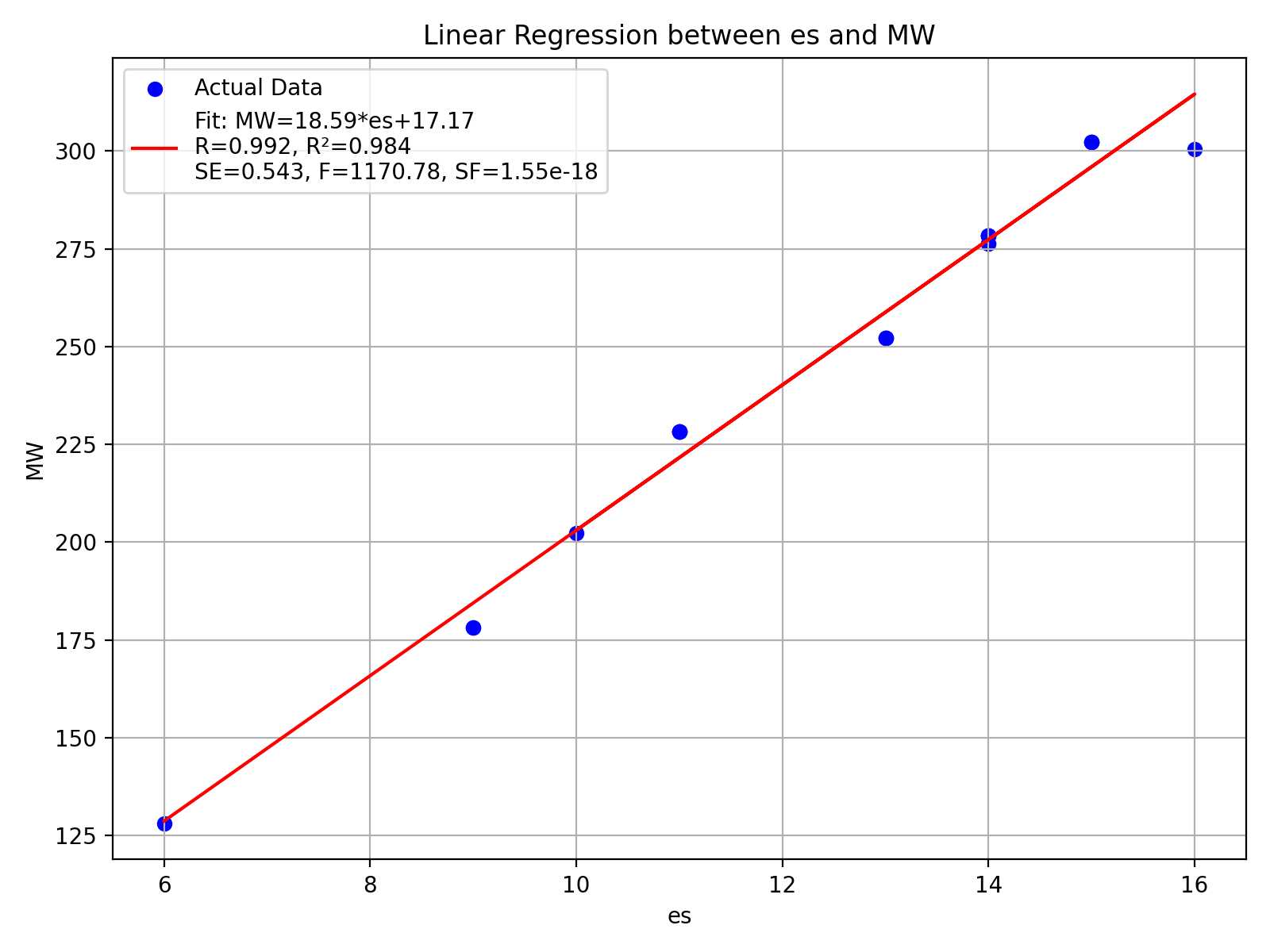}
	\caption{Scatter plot of the linear model between the edge irregularity strength and the molar weight property of benzenoid hydrocarbons}
\end{figure} 

\begin{figure}[hbt!]
	\centering
	\includegraphics[width=120mm]{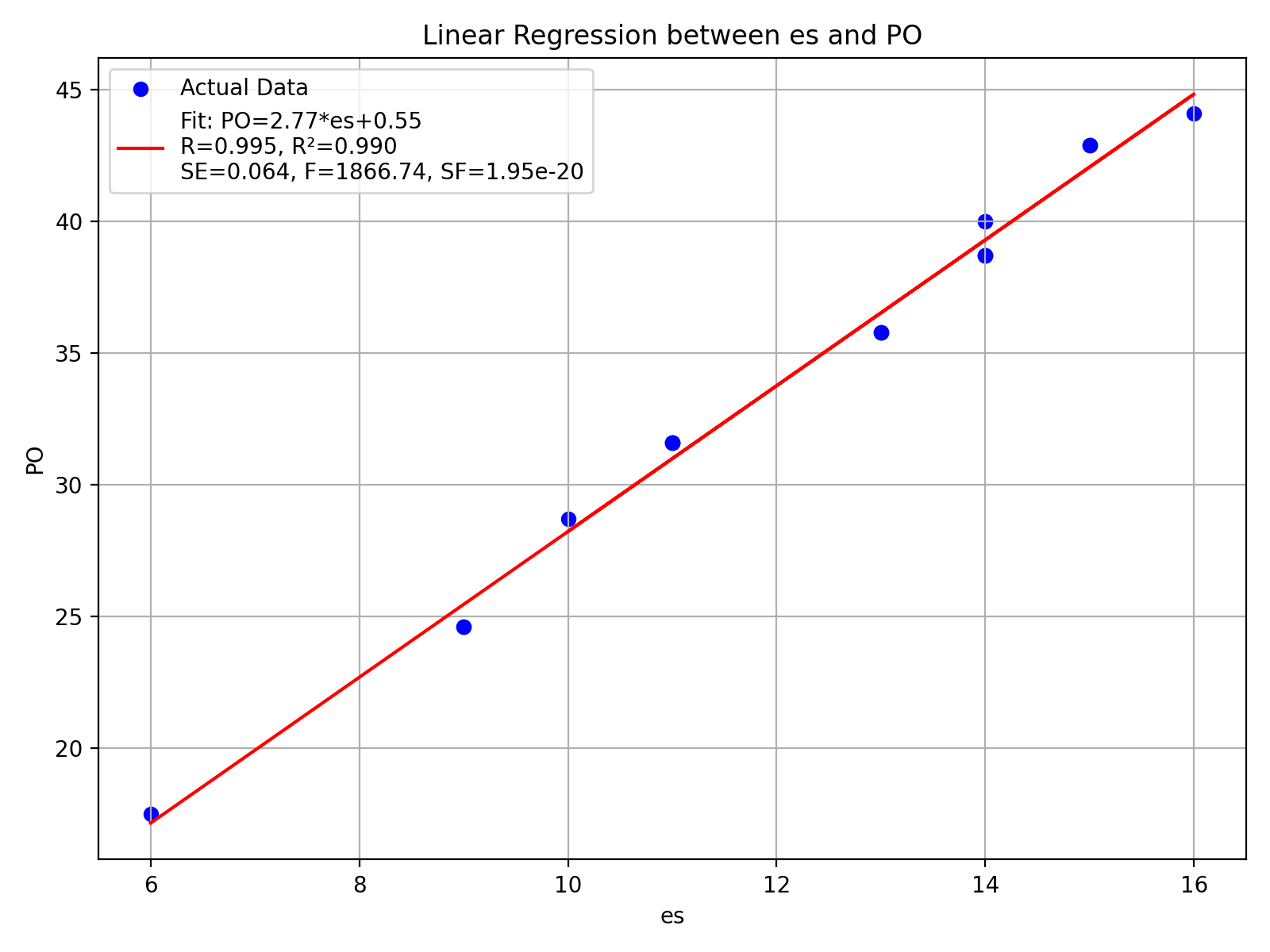}
	\caption{Scatter plot of the linear model between the edge irregularity strength and the polarizability property of benzenoid hydrocarbons}
\end{figure} 

\newpage

\begin{figure}[hbt!]
	\centering
	\includegraphics[width=120mm]{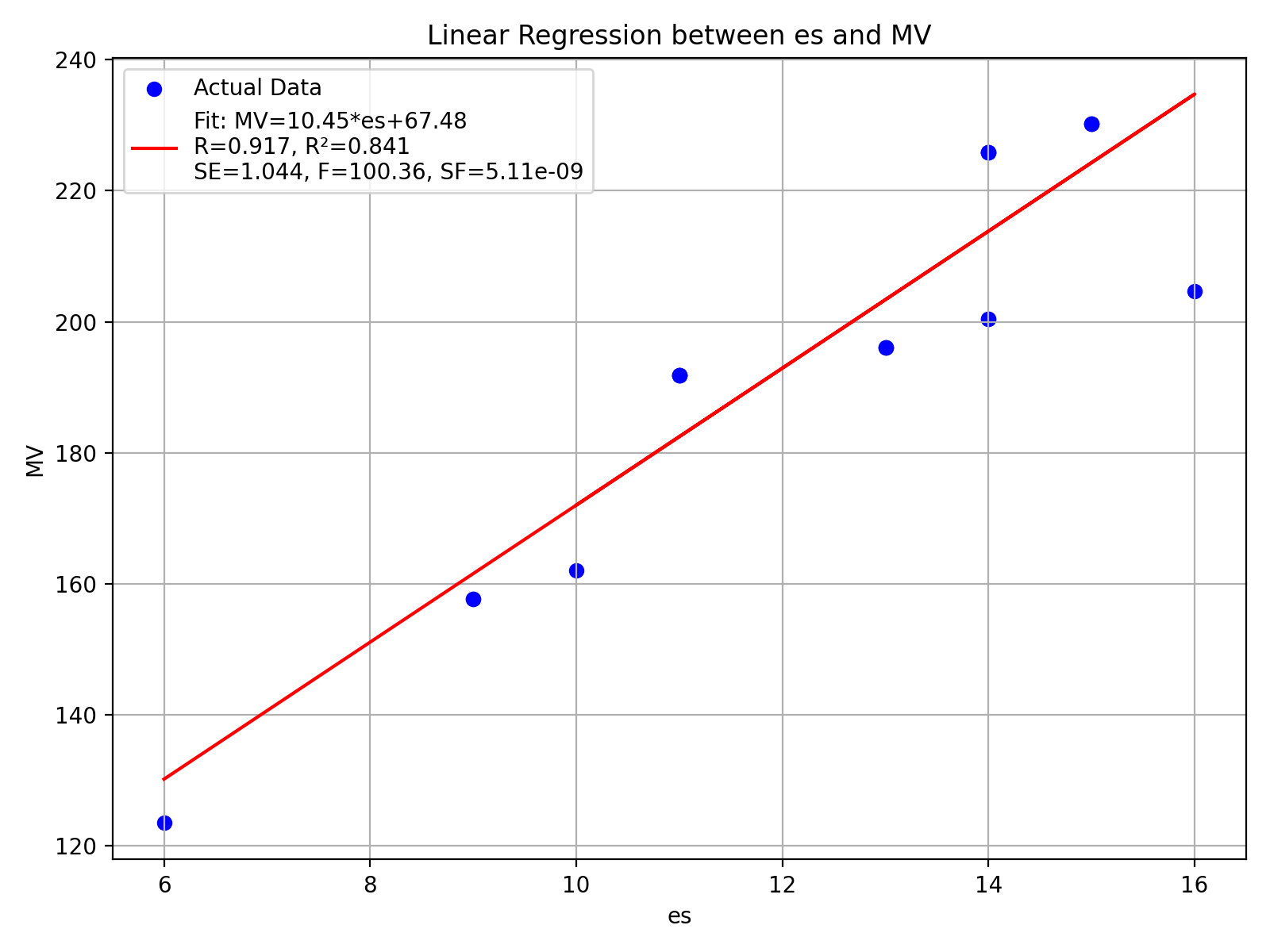}
	\caption{Scatter plot of the linear model between the edge irregularity strength and the molar volume property of benzenoid hydrocarbons}
\end{figure} 

\newpage
\begin{figure}[hbt!]
	\centering
	\includegraphics[width=120mm]{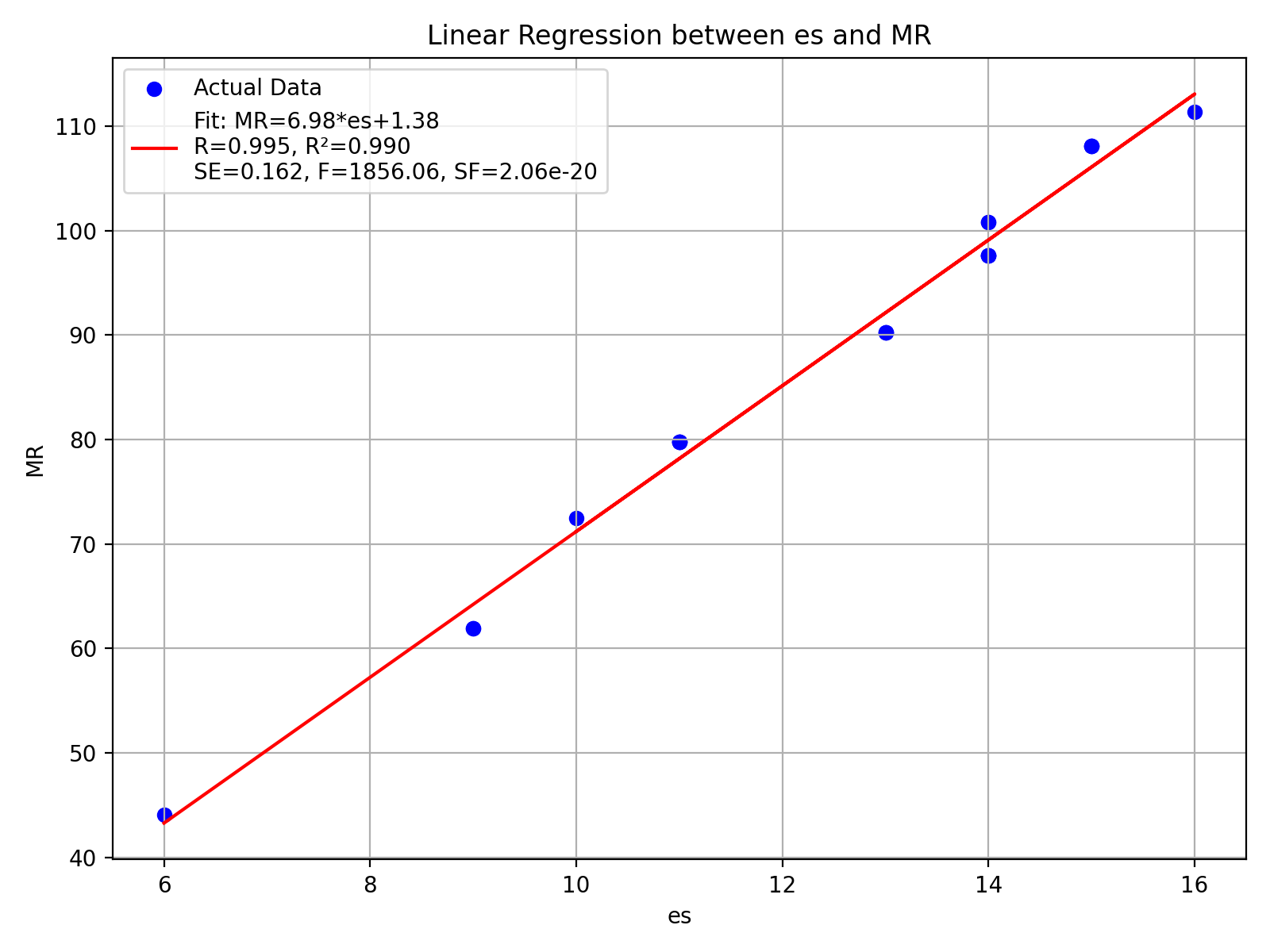}
	\caption{Scatter plot of the linear model between the edge irregularity strength and the molar refractivity property of benzenoid hydrocarbons}
\end{figure} 

\begin{figure}[hbt!]
	\centering
	\includegraphics[width=120mm]{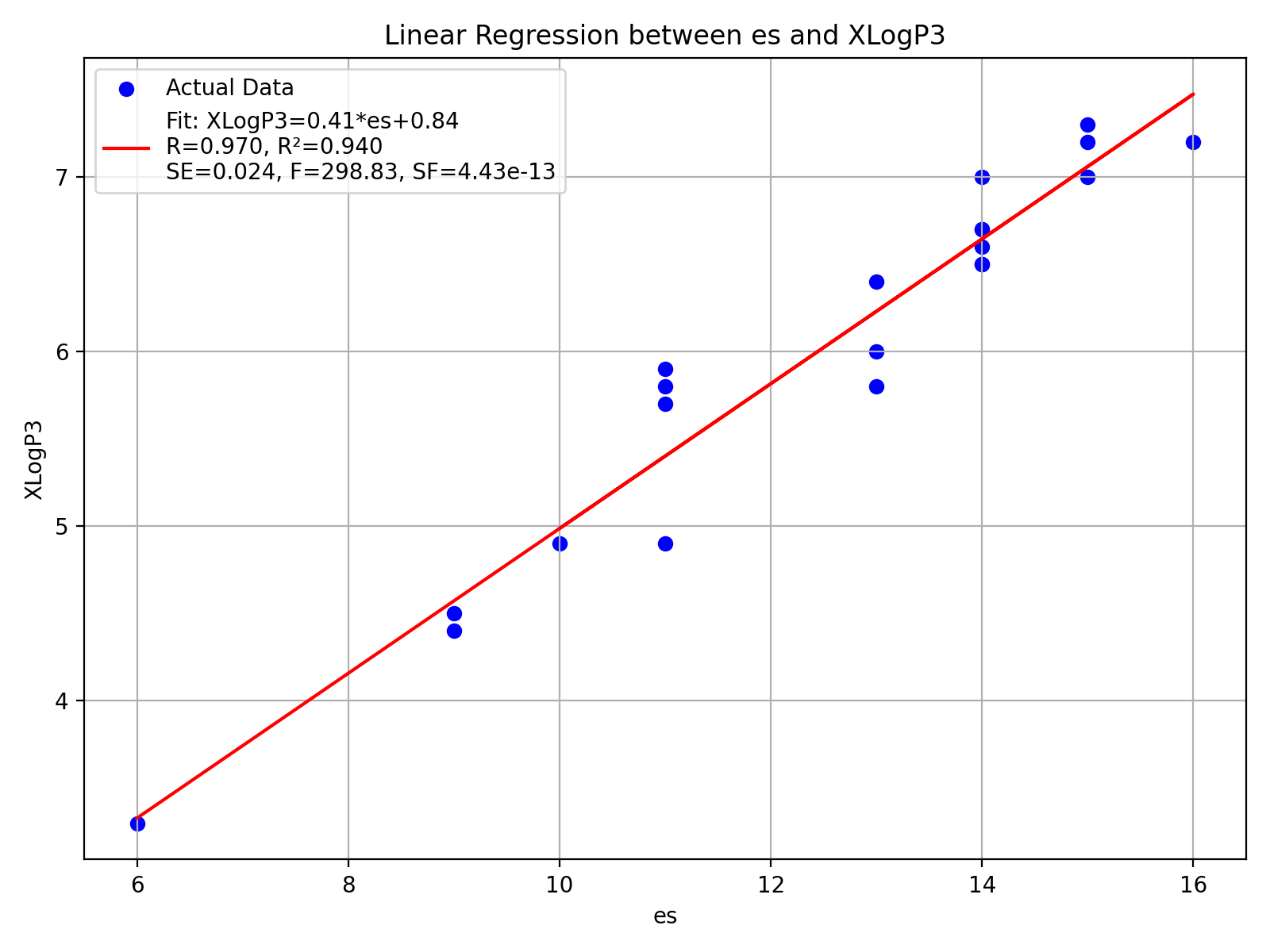}
	\caption{Scatter plot of the linear model between the edge irregularity strength and the XLogP3 property of benzenoid hydrocarbons}
\end{figure} 

\newpage

\begin{figure}[hbt!]
	\centering
	\includegraphics[width=120mm]{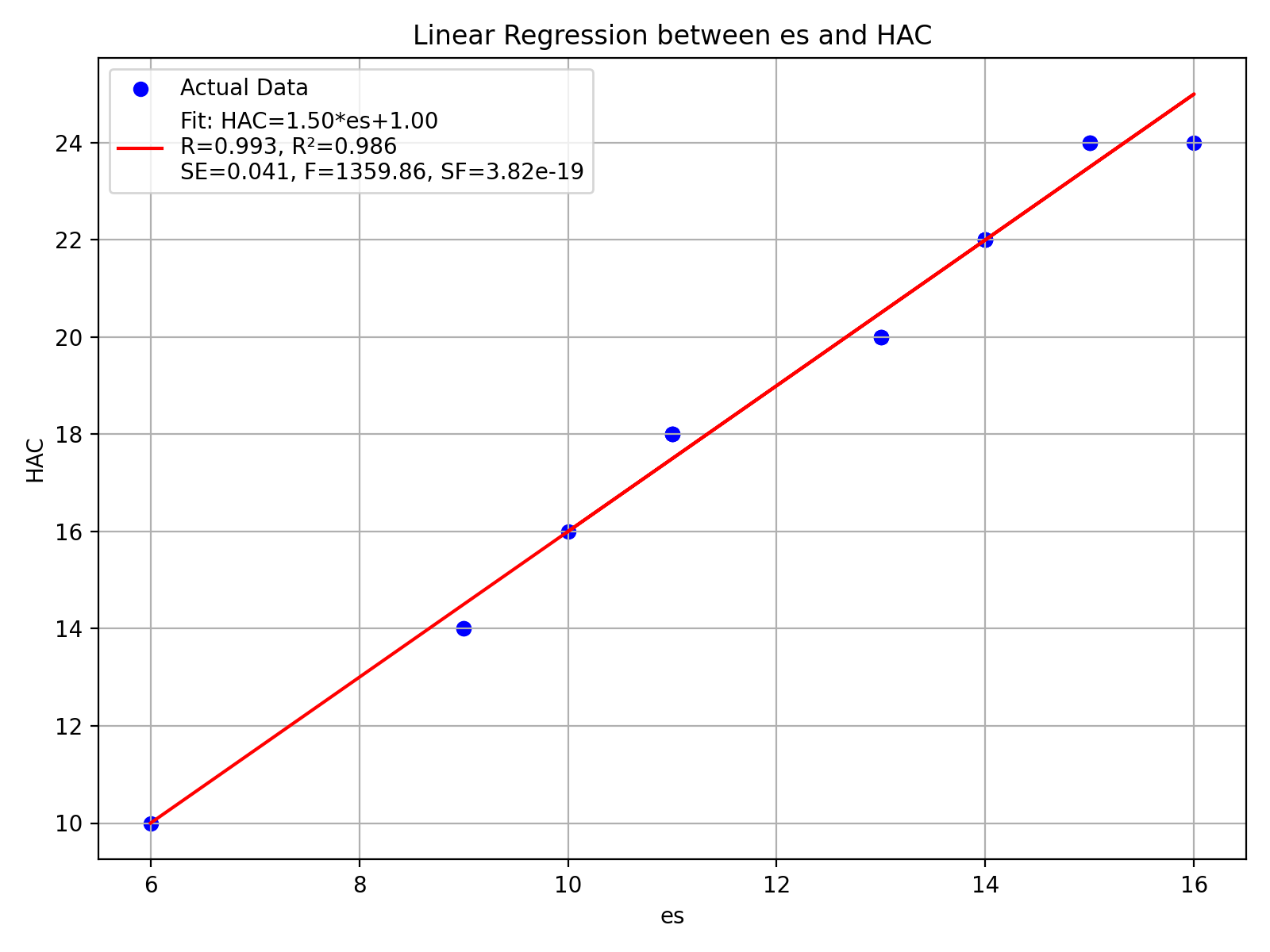}
	\caption{Scatter plot of the linear model between the edge irregularity strength and the heavy-atom count property of benzenoid hydrocarbons}
\end{figure} 

\begin{figure}[hbt!]
	\centering
	\includegraphics[width=120mm]{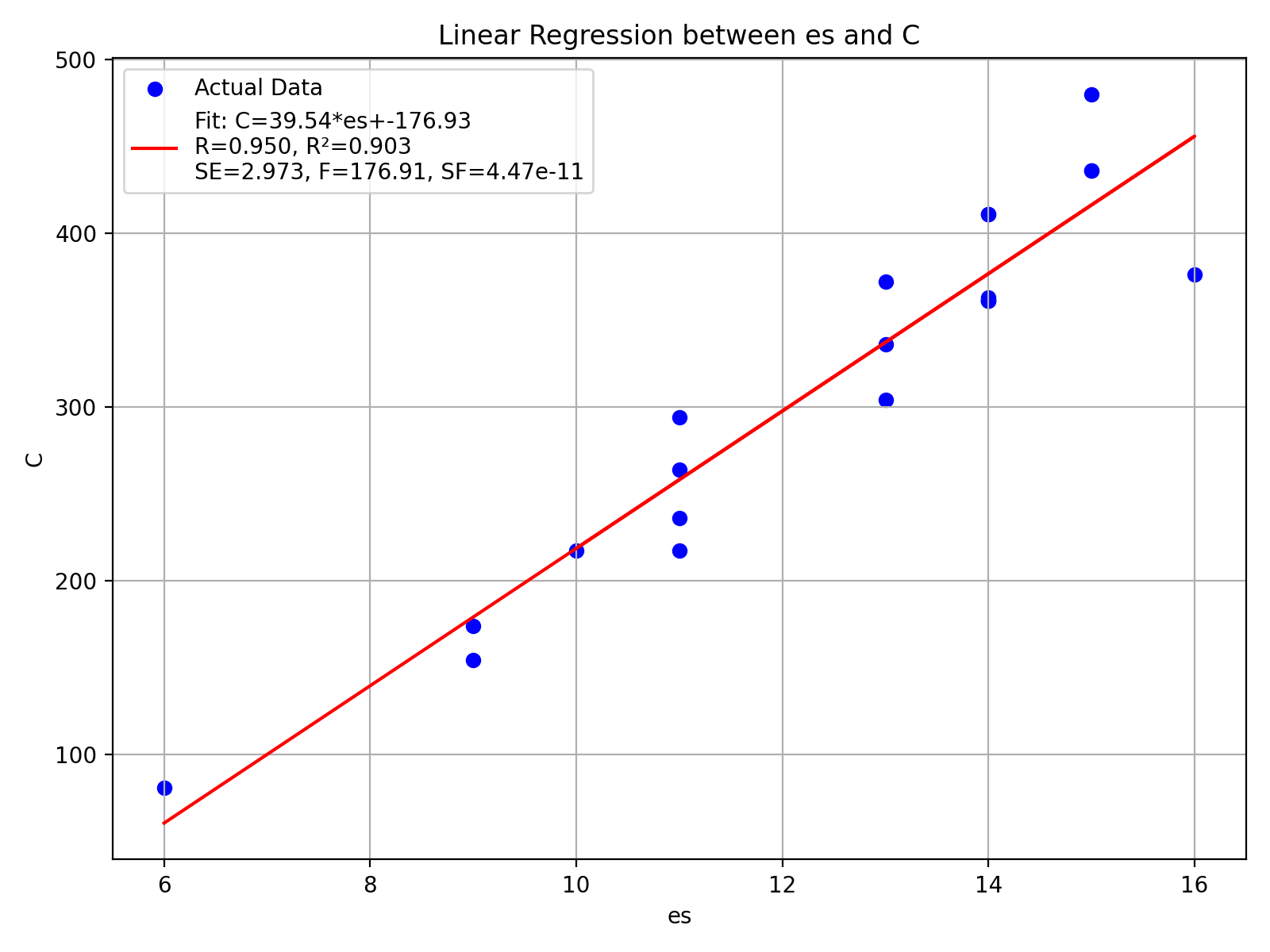}
	\caption{Scatter plot of the linear model between the edge irregularity strength and the complexity property of benzenoid hydrocarbons}
\end{figure} 

\newpage

\section{Comparative study} 

A comparative analysis of the regression results obtained in the present study with those reported in \cite{16} reveals notable differences in the strength of correlations between the edge irregularity strength and various physicochemical properties of benzenoid hydrocarbons.

In \cite{16}, $R^2$ values for most physicochemical properties were very high, ranging from 0.9006 (for MV) to 0.9998 (for PO), indicating an almost perfect linear relationship between the studied parameters and the considered topological index. The present study, however, shows slightly lower but still strong correlations, with $R^2$ values ranging from 0.8408 (for MV) to 0.9899 (for PO and MR). This suggests that while the linear relationships remain statistically significant, the correlations are marginally weaker compared to those in \cite{16}. Specifically, the $R^2$ values for $E_{\pi}$ (0.9892), PO (0.9899), and MR (0.9899) in the present analysis closely approach those from the earlier study (0.9986–0.9998), indicating excellent model reliability and reproducibility across datasets. In contrast, BP and MW exhibit slightly reduced coefficients of determination (0.9437 and 0.9840, respectively) compared to \cite{16} ($R^2 = 0.9558$ and $0.9949$). The parameters MV and C exhibit the largest deviations, with $R^2$ values of 0.8408 and 0.9030 in the current study, compared to 0.9006 and 0.9394 in the earlier work, respectively. These reductions could stem from differences in dataset composition, edge irregularity definitions, or computational models used to estimate physicochemical parameters. Whether we use the topological index from \cite{16} or the edge‐irregularity strength developed here, the results reveal a clear linear relationship with the physicochemical properties of benzenoid hydrocarbons.

\section{Conclusion}
We have shown that edge irregularity strength—a systematic labeling method—matches conventional topological indices in predicting physicochemical properties of benzenoid hydrocarbons, including boiling point (BP), $\pi$‐electron energy ($E_{\pi}$), molecular weight (MW), polarizability (PO), molar volume (MV), molar refractivity (MR), XLogP3, heavy atom count (HAC) and complexity (C). Its rigorous assignment rules capture essential structural information, and our findings demonstrate that edge irregularity strength can serve as an effective alternative or complement to established indices in QSPR/QSAR modeling.

\textbf{The potential limitations of the work:} We see two main limitations in our work. 
\begin{enumerate}
\item Even for smaller or simpler structures, computing the exact edge irregularity strength relies on the detailed steps in Theorem 2.1, which can still be quite involved. 
\item Finding the exact edge irregularity strength for large structures can be difficult, because the number of possible label assignments grows very quickly with size.   
\end{enumerate}

Future work should aim to create faster methods for calculating edge irregularity strength so it can be used more easily in QSPR/QSAR studies.

\section*{Funding} No funding is available for this study.
\section*{Author contributions}  U. Vijaya Chandra Kumar, Narahari N: Conceptualization, methodology; H.M. Nagesh, Conceptualization, methodology, and original draft writing. 
\section*{Data Availability Statement}
The data used support the work cited within the text as references.
\section*{Declarations}
\textbf{Ethical Approval} Not applicable.\\
\textbf{Conflict of interests} The authors declare that they have no known competing financial interests or personal relationships that could have appeared to influence the work reported in this paper.

\end{document}